\documentclass{elsart}

\usepackage{amsmath,amssymb}
\usepackage[latin1]{inputenc}
\usepackage{latexsym}
\usepackage{graphicx}

\newcommand{\eq}[1]{eq.~(\ref{#1})}

\newcommand{\fig}[1]{Figure~\ref{#1}}
\newcommand{\tab}[1]{Table~\ref{#1}}


\begin{document}

\begin{frontmatter}

\title{Fast Parallel I/O on Cluster Computers}

\author[W]{Thomas Düssel},
\author[W,P]{Norbert Eicker},
\author[K]{Florin Isaila},
\author[W]{Thomas Lippert},
\author[W,K]{Thomas Moschny}, 
\author[B]{Hartmut Neff}, 
\author[W]{Klaus Schilling},
\and\author[K]{Walter Tichy}
\address[W]{Department of Physics, University of Wuppertal, Gaußstraße~20, 
            42097~Wuppertal, Germany}
\address[P]{ParTec AG,  Possartstr.~20, 81679~München, Germany}
\address[K]{Institute for Program Structures and 
  Data Organization (IPD), 
  University of Karlsruhe, Postfach~6980, 
  76128~Karlsruhe, Germany}
\address[B]{Physics Department, Boston University, 
590~Commonwealth Avenue Boston,\\ MA~02215, USA}

\begin{abstract}
  Today's cluster computers suffer from slow I/O, which slows down
  I/O-intensive applications. We show that fast disk I/O can be
  achieved by operating a parallel file system over fast networks such
  as Myrinet or Gigabit Ethernet.
  
  In this paper, we demonstrate how the ParaStation3 communication
  system helps speed-up the performance of parallel I/O on clusters
  using the open source parallel virtual file system (PVFS) as testbed
  and production system.  We will describe the set-up of PVFS on the
  Alpha-Linux-Cluster-Engine (ALiCE) located at Wuppertal University,
  Germany.  Benchmarks on ALiCE achieve write-performances of up to
  1~GB/s from a 32-processor compute-partition to a 32-processor PVFS
  I/O-partition, outperforming known benchmark results for PVFS on the
  same network by more than a factor of 2.  Read-performance from
  buffer-cache reaches up to 2.2~GB/s.  Our benchmarks are giant,
  I/O-intensive eigenmode problems from lattice quantum
  chromodynamics, demonstrating stability and performance of PVFS over
  Parastation in large-scale production runs.\\[6pt]

\end{abstract}

\end{frontmatter}

\section{Introduction}

Within the past years commodity-off-the-shelf (COTS) clusters have
evolved towards cost-effective general-purpose HPC devices.  Such
systems, self-made and commonly denoted as Beowulf
computers~\cite{ridge97beowulf}, show up increasingly on the TOP500
list~\cite{top500}.  While gigabit network technology
(Gigabit-Ethernet, Myrinet) along with error-correcting
zero-copy-communication software~\cite{GM,SCORE,PARTEC} have boosted
communication-intensive number-crunching tasks, I/O-intensive
computations have not benefited from cluster computers to the same
extent.  The reason was the relatively slow I/O capability of
clusters.

Today, a promising approach to achieve fast I/O on cluster computers
is to utilize distributed disks and their aggregate bandwidth by means
of a parallel file system (PFS).  A PFS is designed to make the entire
disk capacity of the I/O-nodes available to all the compute-nodes and
to allow the parallel file access of the compute nodes to be
translated into real parallel disk access.  Physically, files are
stored on a given partition of cluster nodes by distributing the data
of the given file, for instance in a round robin fashion.  In contrast
to standard network file systems, a PFS provides concurrent parallel
access to store or read the file from all nodes of a parallel
application.  In a typical implementation, a set of compute-nodes
reads and writes data to another set of I/O-nodes that host the
physical resources of the PFS\@.  The two sets of nodes may be
identical, may partly overlap or may even be distinct.  In principle
this concept allows for scalability of the I/O-rate with the number of
I/O-nodes, provided that \emph{(i)} the number of compute-nodes is
large enough to saturate the capacity of the I/O-nodes---this is
usually the case as soon as the number of compute-nodes equals the
number of I/O-nodes---and \emph{(ii)} the network delivers full
bi-sectional bandwidth---this is, for instance, the case for crossbar
or multi-stage crossbar topologies.

Distributed file systems like NFS or AFS are not suited for concurrent
high-bandwidth file-access as required in I/O-intensive cluster
computing applications.  In these systems, parallel data accesses of
compute nodes are serialized by file servers. Therefore, they can not
be called "parallel" file systems.  There exist commercially available
parallel file systems (sometimes platform dependent), for example the
General Parallel File System GPFS (IBM~\cite{GPFS}).  Currently, the
only known open source parallel file system, working in a stable
manner, and freely available for Linux under the GNU General Public
License, is the Parallel Virtual File System (PVFS) developed at
Clemson University and Argonne National Laboratory~\cite{PVFS-Site}.
PVFS is devised as a truly parallel file system for use on cluster
computers. The communication back-end of the standard distribution of
PVFS is based on the TCP/IP protocol.  Therefore, PVFS can readily be
operated on top of any network supporting TCP/IP\@.

In this paper we consider PVFS boosted by the Myrinet communication
network~\cite{MYRINET} of the Alpha Linux Cluster Engine (ALiCE)
located at Wuppertal University, Germany~\cite{PIK:2002}.  There are
several error-correcting and package-loss-safe communication
sub-systems available, designed to drive Myrinet: e.g.\ the vendor-provided GM
software~\cite{MYRINET}, SCore~\cite{SCORE}, and the ParaStation
system~\cite{PARTEC}, developed at Karlsruhe University.  On ALiCE, we
are using ParaStation.

ParaStation implements the concept of virtual nodes, operating in
close interaction with queuing systems like PBS~\cite{PBS}.  The
communication system provides safe multi-user operation and
outstanding stability, not to mention the comfortable
single-point-of-administration management by means of the
ParaStation-daemons. System crashes are tidily cleaned-up without any
user interference. Most important in our context is however the
communication bandwidth under ParaStation.  A special kernel module
routes TCP/IP via the ParaStation communication system and renders
Myrinet an additional IP-network with full bi-sectional bandwidth. In
this manner, the superior bandwidth from ParaStation as Myrinet driver
can be exploited.

Combining parallel file systems like PVFS with ParaStation perfectly
meets the demands of an application from the post-simulation phase of
a large scale Monte Carlo project in lattice-quantum chromodynamics
(LQCD).  We are evaluating giant eigenproblems~\cite{Neff:2001zr},
which are very data-intensive, on cluster computers. The eigenvectors
are required for the construction of correlations between two quark
loops. Such creation and annihilation of quarks originate from the
quantum fluctuations of the QCD vacuum, according to Heisenberg's
uncertainty principle. They are considered affectual to the unusually
large mass of the $\eta'$-meson~\cite{Schilling:2002gm}.  In our
simulations, we have to compute $\mathcal O(1000)$ low eigenvectors of
the fermionic matrix, which describes the dynamics of the quarks.  The
size of each vector is $\mathcal O(10^6)$.

Typically, about 10~GB of I/O is carried out in data-intensive
production steps of about 10~minutes compute time on 16 to
64~processors; actually several thousands of such runs are performed.
Without parallel I/O, reading or writing lasts between 10 and
30~minutes.  PVFS helps cut down the read and write times to about
20~seconds.

The paper is organized as follows: in Section~\ref{SETUP} we present
the new TCP/IP kernel module included in ParaStation and describe the
connectivity of ALiCE and its specific PVFS implementation.
Section~\ref{SPEEDS} gives benchmarks for the components of the
I/O-machinery, including disk-speed and TCP/IP node-to-node rates.  The
results of the PVFS benchmarks are shown in Section~\ref{PVFSBENCH}
along with a comparison with Ref.~\cite{PVFSpaper}.  We are in the
position to test PVFS/ParaStation within a large-scale application
from lattice quantum chromodynamics, described in Section~\ref{EIGEN}.
Finally we summarize and give a short outlook on the novel parallel
file system ClusterFile that is currently under development at the
IPD, University of Karlsruhe~\cite{IT01}.

\newpage

\section{Technical Background\label{SETUP}}

For cluster applications based on IP-communication to benefit from
ParaStation's performance, a TCP/IP kernel module has been introduced
on top of the ParaStation communication system.  In this manner,
stable communication with gigabyte bandwidth is provided for the
parallel virtual file system (PVFS) as implemented on the 128-node
Alpha-Linux-Cluster-Engine ALiCE\@.

\subsection{TCP/IP kernel module with ParaStation\label{MODULE}}

Many cluster applications neither use MPI nor the low-level
communication API as for instance provided by ParaStation.  To benefit
from ParaStation's performance for such applications, an additional
TCP/IP module was developed at the Institute for Programs and Data
Structures, University of Karlsruhe.

This module provides a network driver interface to the Linux kernel,
just as any other Ethernet card driver does. This way, virtually any
Ethernet protocol that is supported by the kernel can be transported
over Myrinet. In practice, most applications will use the TCP (or UDP)
over IP protocol\footnote{The intended use is the reason for the
  somewhat misleading name of the module. It should better be called
  Ethernet driver for ParaStation.}. 

Internally, the module uses the kernel variant of the ParaStation
low-level communication interface to send raw Ethernet datagrams to
any other host in the cluster. This interface supports a set of basic
communication operations, since at this level  we don't need more
functionality. Packet (dis)assembling for instance is done by the
kernel. Upon startup, a so-called \textit{kernel-context} is obtained
from the ParaStation module. This reserves a certain number of
communication buffers in the memory of the Myrinet network adapter
card and instructs the driver to listen for messages addressed to the
TCP/IP module.  Secondly, a call-back mechanism is established:
whenever such a message arrives, an interrupt is risen and the
ParaStation interrupt handler calls a method of the TCP/IP module
which in turn hands the message to the Linux network stack. If a
message is to be sent, the network stack functions call a method of
the module that was registered upon initialization, handing the
message over to the ParaStation module.

In order to address other nodes in the cluster, the TCP/IP driver
module (in fact an Ethernet driver) maps the ParaStation node
identification number (ParaStation ID) onto Ethernet hardware
addresses. The administrator can set up static ARP tables that map
IP addresses to these hardware addresses. This is necessary, as the
driver does not support broadcast messages that would be required to
enable automatic ARP functionality. However, if IP addresses are
chosen such that the ParaStation ID (the node number counted from 0)
equals the IP address minus one, the static ARP tables can be omitted.
In this case, any driver module in the cluster can guess ParaStation
IDs from IP addresses and thus generate fake ARP reply messages.

\subsection{ALiCE setup}

The Alpha-Linux-Cluster-Engine ALiCE is an assembly of 128 Compaq DS10
workstations~\cite{PIK:2002}. ALiCE, located at the University of
Wuppertal (Germany), is fully operational since the end of 2000. The
machine is equipped with Alpha 21264 EV67 processors, with 2~MB cache,
operating at a frequency of 616~MHz. With 256~MB ECC memory for each
processor, the total amount of memory is 32~GB\@. The disk space of
10~GB per node adds up to about 1.3~TB in total.  The nodes are
interconnected by a 2 $\times$ 1.28~Gbit/s Myrinet network configured
as a multi-stage crossbar with full bi-sectional bandwidth.  The
M2M-PCI64A-2 Myrinet cards utilize a 64~bit/33~MHz PCI bus.

\begin{figure}[!htb]
\begin{center}
  \includegraphics[width=.82\textwidth]{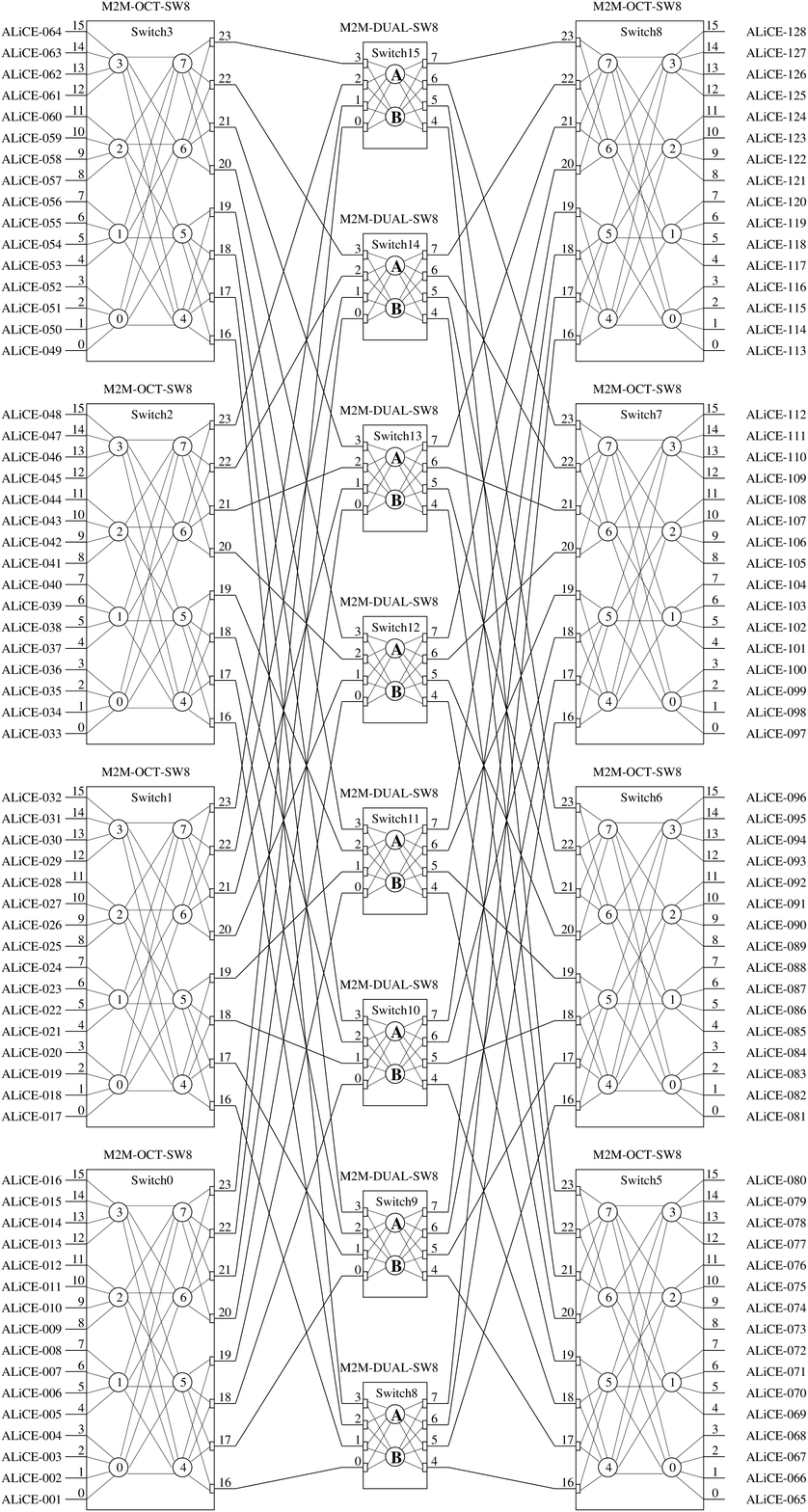}
\end{center}
\caption{Myrinet multi-stage crossbar network.\label{MYRINET}}
\end{figure}

\fig{MYRINET} shows the hardware plan of the inter-node connection of
ALiCE\@. A full crossbar is realized by three switch-stages (where
each octal switch actually consists of two stages) employing the
$2\times 4$ Myrinet M2M-oct-SW8 switches and 8~Myrinet M2M-dual-SW8
switches. The hardware latency is about 100~ns per switch stage, far
below the total latency (software and hardware) of 17.1~$\mu$s.

\begin{figure}[!htb]
\begin{center}
\includegraphics[width=\textwidth]{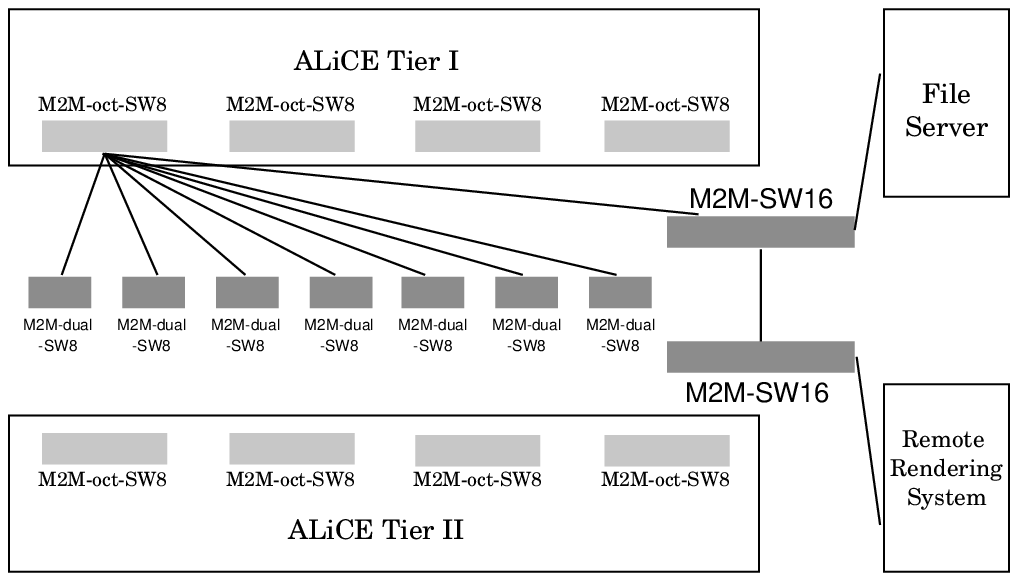}
\end{center}
\caption{Inclusion of external devices (not all connections 
  drawn).\label{MYRINET2}}
\end{figure}

We have extended the network in order to incorporate an external file
and archive server as well as to provide gigabit links to external
machines for the purpose of fast on-line visualization.  To this end,
we have exchanged two of the 8 M2M-dual-SW8 switches by two M2M-SW16
switches, as sketched in \fig{MYRINET2}.  This way, we could avoid an
inhomogeneous number of hardware stages of the multi-stage crossbar
network.

\subsection{PVFS partitioning}

\begin{figure}[!htb]
\begin{center}
\includegraphics[width=.8\textwidth]{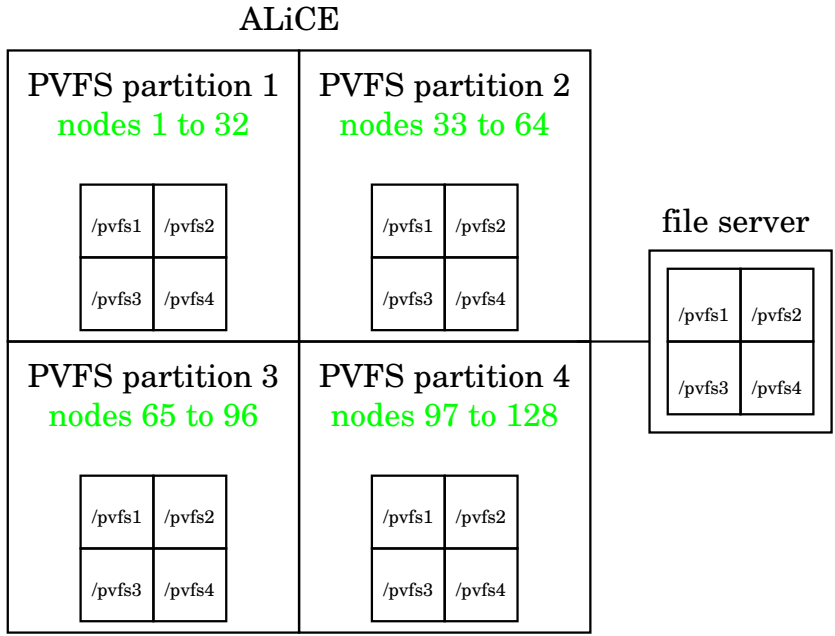}
\end{center}
\caption{Organization of 4 PVFS partitions on ALiCE.\label{SETUPPVFS}}
\end{figure}

On ALiCE, we run 4 different PVFS partitions with 32-nodes each. This
partitioning fits well with the compute-partitions, chosen such as to
optimize the compute performances of our applications. Each PVFS
partition (\texttt{/pvfs1} to \texttt{/pvfs4}) is represented by a
mount point on each node and on the file server.  Mounting PVFS on the
file server enables us to copy UNIX files with ParaStation TCP/IP
speed from the external RAID to PVFS\@.  For each partition, the last
node plays the rôle of the management node.  The entire set-up is
displayed in \fig{SETUPPVFS}.

\section{Performance of Basic I/O-Devices\label{SPEEDS}}

The proper interpretation of the benchmark results presented in
Section~\ref{PVFSBENCH} requires some background knowledge about the
features of ALiCE's basic I/O-components, i.e., file system
performance, TCP/IP and MPI data rates.

\subsection{Local file system performances\label{DIRTY}}

\begin{figure}[!htb]
\begin{center}
\includegraphics[width=.8\textwidth]{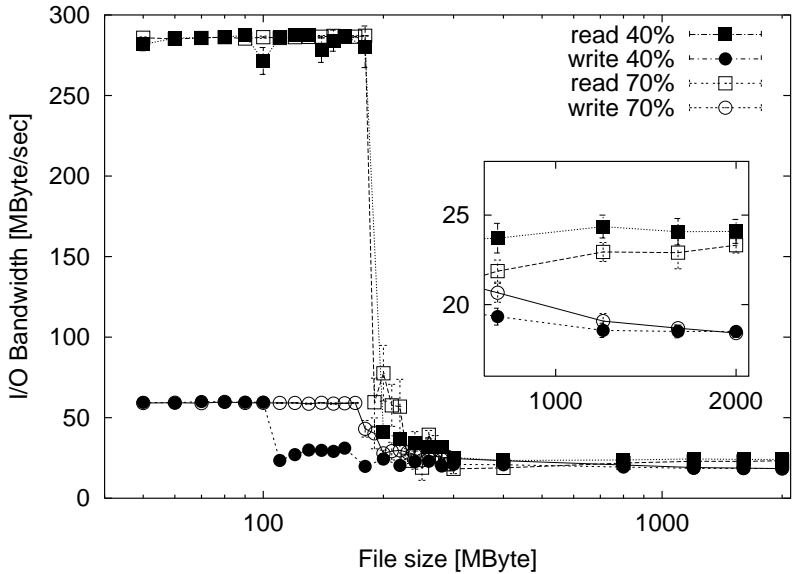}
\end{center}
\caption{Local file system performances on ALiCE.\label{BONNIE}}
\end{figure}

The ALiCE DS10 nodes are equipped with 10~GB Maxtor IDE
disks\footnote{5400~rpm versions.}.  We have carried out tests on
local file systems, formatted with ReiserFS, by means of
\texttt{bonnie++}~\cite{BONNIE}.  Write and read-performances are
determined for a series of file sizes, from 10~MB up to 2~GB\@.  In
order to reveal buffer-cache effects, we have adjusted the ``dirty
buffer'' parameter\footnote{The standard behavior of the Linux kernel
  is to start flushing buffer pages as soon as the given percentage of
  memory available for buffer cache is ``dirty''. A buffer page is
  called dirty if it changed in memory but has not yet been written to
  disk.} to two different values, 40~\% and 70~\%. Our results are
given in \fig{BONNIE}.

The \texttt{bonnie++} benchmark works as follows: the given file is
written and read back immediately.  We observe the write performance
to buffer cache of 60~MB/s to fall to about 23~MB/s at a file size of
100 and~200 MB, for 40 and 70~\% dirty buffers, respectively.
Asymptotically, the performance is decreasing to about 19~MB/s.  Being
able to read from the buffer cache, the subsequent read operation
shows a bandwidth of more than 280~MB/s for small files.  At a file
size of about 200~MB, the speed drops down to 24~MB/s.

\subsection{TCP/IP performance via ParaStation}

\begin{figure}[htb]
\begin{center}
  \includegraphics[width=.8\textwidth]{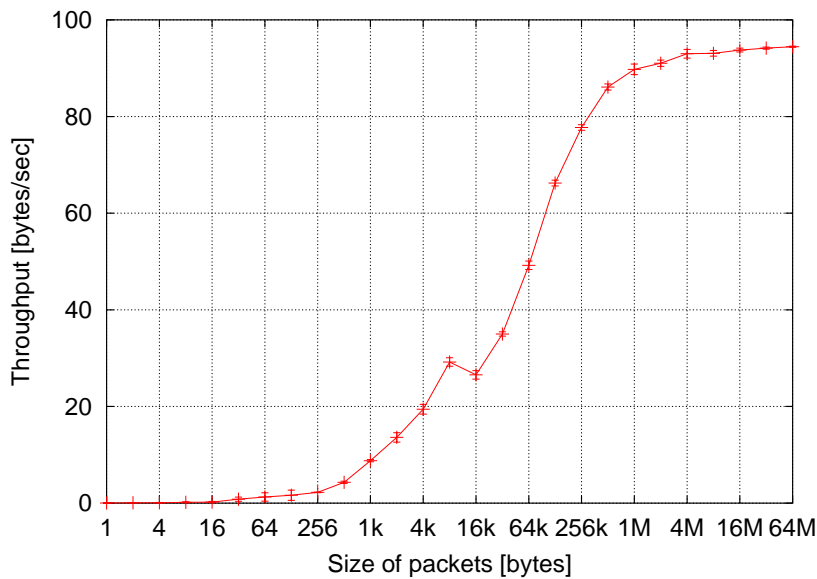}
\end{center}
\caption{Performance test by \texttt{ttcp} via ParaStation.\label{TCPFIG}}
\end{figure}

On ALiCE, we can choose TCP packages to be routed via Fast Ethernet or
alternatively---using the new TCP/IP kernel module described in
Section~\ref{MODULE}---via ParaStation/Myrinet.  The \texttt{ttcp}
benchmark issues TCP/IP packets over a point-to-point connection to
determine the uni-directional TCP/IP speed, cf.\ \fig{TCPFIG}.  The
performance is seen to saturate at 93~MB/s.

\begin{figure}[!htb]
\begin{center}
  \includegraphics[width=.7\textwidth]{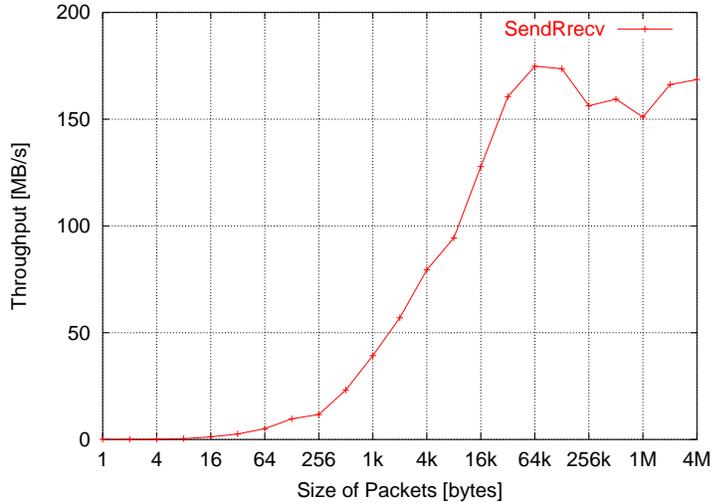}
\end{center}
 \caption{Performances of the 
   \texttt{send-receive} Pallas MPI-benchmark on ALiCE.\label{PALLAS}}
\end{figure}

It is instructive to compare these results with the outcome of the
Pallas \texttt{send-receive} MPI benchmark~\cite{PALLASMPI}, see
\fig{PALLAS}.

\begin{figure}[htb]
\begin{center}
  \includegraphics[width=.7\textwidth]{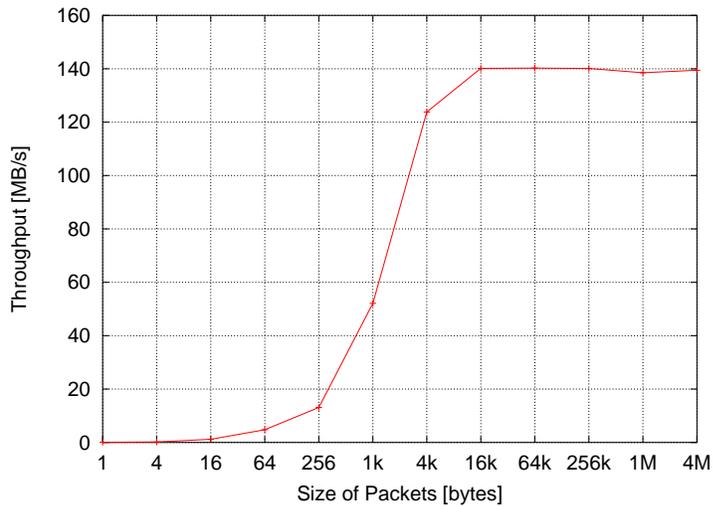}
\end{center}
 \caption{MPI-performance for uni-directional communication.\label{MPI_UNIDIR}}
\end{figure}
 
For the \texttt{send-receive} case (i.e.\ the bi-directional
situation), the performance levels off at a total bandwidth of about
175~MB/s (adding up the data rates of both directions).  As the PALLAS
benchmark does not provide uni-directional measurements, we have
prepared corresponding \texttt{send} and \texttt{receive} programs and
found a saturation of the uni-directional MPI-performance at about
140~MB/s, cf.\ \fig{MPI_UNIDIR}.  The data rate via TCP/IP is only
about 34~\% smaller than via MPI, in spite of the overheads of the
full-fledged TCP/IP implementation.  Still this leaves room for
improvement, since on a cluster there exists a priori knowledge of the
paths to all IP destinations, therefore one could try to set up a slim
TCP/IP protocol.  Moreover, a double error checking is carried out,
one on the ParaStation level, and a second one on the TCP/IP level.

\section{File System Benchmarks\label{PVFSBENCH}}

Our benchmarks follow Ref.~\cite{PVFSpaper}, where performance results
on 60~nodes of the Chiba-City cluster at Argonne National Laboratory
have been reported.  This cluster is equipped with the same Myrinet
version as ALiCE\@. This will enable us to compare our results using
TCP/IP over ParaStation on an Alpha system with TCP/IP over the
Myrinet/GM software on a Pentium cluster.  Since loading and
discharging large amounts of data to the cluster constitute a crucial
bottleneck for data-intensive production runs on clusters, we include
performances with respect to reading from and writing to UNIX files
located on an external RAID\@.

\subsection{Concurrent read/write performance}

Our test code works as follows: a new PVFS file, common to $P$
compute processes, is opened on $N$ I/O-nodes. Concurrently the same
amount of data $S$ is written from each of the $P$ processes (virtual
partitioning) to disjoint parts of the file. PVFS stripes
the data onto the $N$ I/O-nodes (physical partitioning)
with a stripe size of 64~kB.

After the data is written, we close the file and reopen it again in
order to reshuffle the same data back to the compute-nodes.  The
bandwidth for write and read operations is computed from the maximum
of the wall clock execution times achieved on all the $P$
compute-nodes.

\begin{figure}[htb]
\begin{center}
  \includegraphics[width=.8\textwidth]{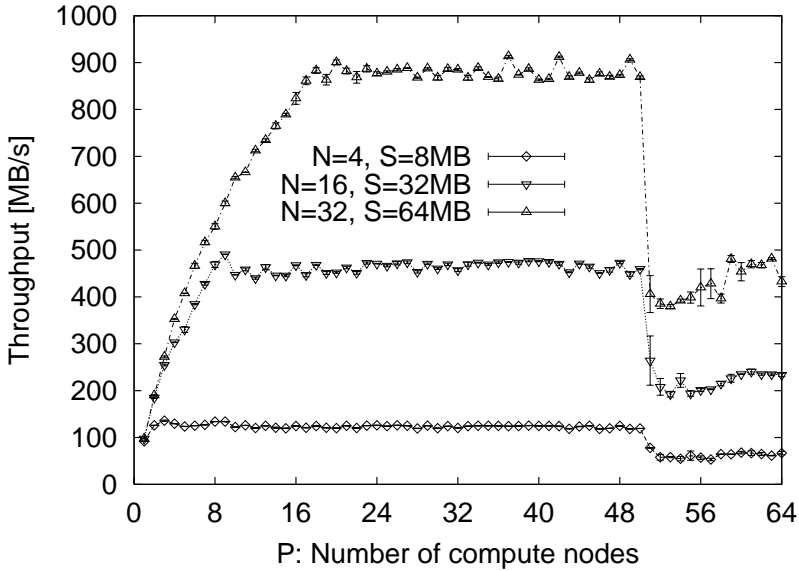}
\end{center}
\caption{Concurrent write performance.\label{WRITE}}
\end{figure}

We vary $P$ in the range $P=1\dots 64$, and repeat each measurement
for $N=4$, $N=16$ and $N=32$ I/O-nodes. The amount of data $S$ written
and read \textit{per compute-node} is chosen proportional to the
number $N$ of I/O-nodes, $S/N = \textrm{const}$ (here we follow
closely the benchmark of Ref.~\cite{PVFSpaper}). The reasoning is that
although we vary the number of I/O-nodes, the buffer-cache will be
saturated for one and the same number of compute-nodes.  Indeed this
behavior is borne out in \fig{WRITE}, which shows the cumulative
throughput for the write operation.

We followed Ref.~\cite{PVFSpaper} and have carried out 5~measurements
in each case.  The smallest and the largest results were discarded and
the remaining ones have been averaged. Actually, the five values did
not differ by more than 2~\% in any of the measurements.

As we see from \fig{WRITE}, the performance quickly reaches a plateau
for each I/O-partition. There is no visible impact from the number of
compute-nodes as long as the buffer-cache of the I/O-nodes is not
saturated. This occurs when the number of compute-nodes is greater
than 50. The amount of data written and read on every I/O-node is
$P\cdot S/N$, so it is greater than 100~MB for $P>50$ and
$S/N=2\,\textrm{MB}$\footnote{In fact the buffer-cache itself is not
  saturated, but at this point the amount of ``dirty buffers'' has
  reached 40~\% of the whole buffer-cache, see Section~\ref{DIRTY}.
  With no other program running at the same time, almost all of the
  main memory is used for the buffer-cache, and 40~\% of 256~MB yield
  100~MB, as observed.}.

The write performance reaches between 29 and 35~MB/s for each
I/O-node, not exhausting the \texttt{bonnie++} figures of
Section~\ref{DIRTY} or the TCP/IP speed, see \fig{TCPFIG}.  However,
we achieve about 30~\% faster write performances than reported in
Ref.~\cite{PVFSpaper}.  Visually comparing the Figure~6 in
Ref.~\cite{PVFSpaper} with \fig{WRITE} we recognize the substantially
improved stability of our curves, a well known feature of the
communication sub-system ParaStation.

After buffer-cache saturation, the performance drops down to a value
which is about 18~\% smaller than expected from the hard disk
performance benchmarks displayed in \fig{BONNIE}.

\begin{figure}[htb]
\begin{center}
  \includegraphics[width=.8\textwidth]{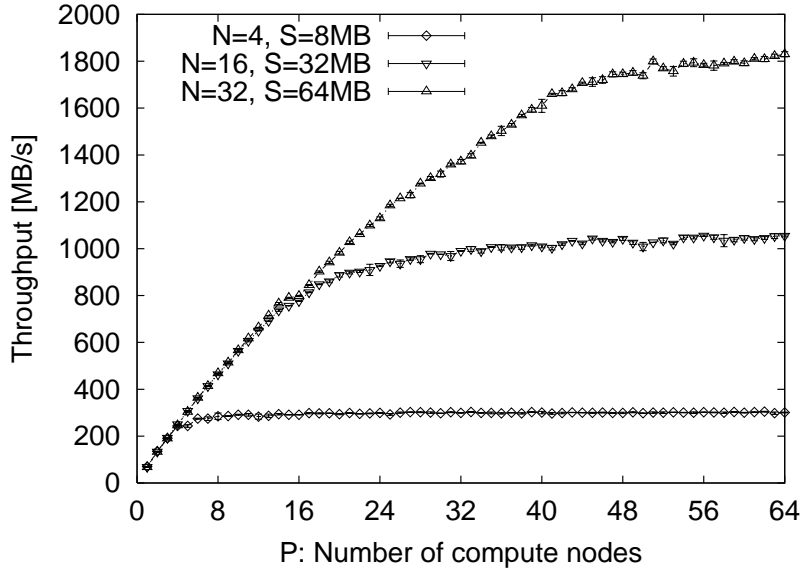}
\end{center}
\caption{Concurrent read performance.\label{READ}}
\end{figure}

As the read operation is carried out directly after the write, with
only a synchronizing barrier in between, the read process can draw the
data directly from buffer-cache. As explained above, dirty buffers are
written to disk if their size exhausts the limit of 100~MB\@. However,
they remain in memory and can be read back at a rate limited only by
the memory bandwidth. Thus, \fig{READ} shows no loss of performance
throughout the test range.

It is gratifying to find that each I/O-node can send with a speed
varying between 56~MB/s and 75~MB/s, since several sockets are served
simultaneously. This performance is just 20~\% slower than the
measured point-to-point performance via TCP/IP, but still about 45~\%
slower than the actual capabilities of ParaStation as seen in MPI
applications, see \fig{PALLAS}.  The maximal performance reaches more
than 1800~MB/s for 32 I/O-nodes.

We should remark, that the read test seems to be rather artificial.
Actually, it would be more meaningful given a hard disk with a read
performance faster than 75~MB/s.  In general, a real application reads
data from disk and not from buffer-cache. In that case, one expects a
saturation below hard disk read performance, as demonstrated in
Section~\ref{EIGEN}.

\begin{figure}[htb]
\begin{center}
  \includegraphics[width=.8\textwidth]{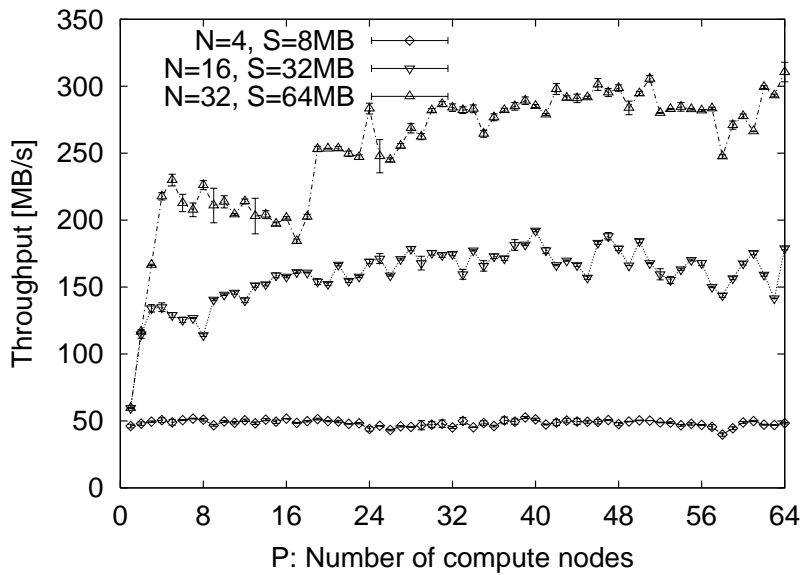}
\end{center}
\caption{Concurrent read performance from disk.\label{READ-DISK}}
\end{figure}

In order to test the raw throughput of the disk, i.e.\ without
utilizing the buffer cache, a modified benchmark was used.
Now, a huge amount of data (multiple times the size of the buffer
cache) is created and written to several files.  After writing and
closing the last file, it is very unlikely that any data from the
first files is still present in the buffer cache.  A subsequent read
operation will therefore be forced to read directly from hard-disk.
\fig{READ-DISK} shows the result of this benchmark.

Obviously the throughput for read operations from PVFS drops
dramatically within such realistic setup. In the case of 4~I/O-nodes
the read performance drops to about 13~MB/s per I/O-node.
Nevertheless a total read performance of 300~MB/s can be achieved if
all 32~I/O-nodes are utilized.

The results presented so far have used a stripe size of 64~kB, the
default value of PVFS\@. Taking a stripe size as large as the amount
of data a given compute-node has to write, we achieve about 920~MB/s
writing from 32 compute-nodes to 32 I/O-nodes. The corresponding
read-operation achieves up to 2200~MB/s using the cache.
Without a full cache, performance is lower.

A second important feature---as far as data-intensive computations on
clusters are concerned---is the speed for charging and discharging the
system. A high throughput is  is crucial for the success of
computer experiments that work on data sets much larger than the PVFS
disk space available. In this case one has to retrieve (store) data
from (to) an external TB-size repository.

Distributing UNIX files onto PVFS from the archive in principle could
proceed with the TCP/IP performance of about 92~MB/s.  Actually, the
limiting factor is the disk performance of the file server. Our RAID
delivers or stores at about 25~MB/s.  In practice this limitation
poses no real problem, since data staging can be carried out
asynchronously to the parallel applications on the cluster.

\section{Fast I/O for Giant Eigenproblems in Lattice Field 
Theory\label{EIGEN}}

In the following, we demonstrate how PVFS/ParaStation enables us to
compute huge eigensystems on cluster computers.  Such computational
problems arise in the post simulation phase of Monte Carlo simulations
of lattice quantum chromodynamics (LQCD).  We aim at computing
$\mathcal O(1000)$ low eigenvectors of the so-called fermionic matrix, which
describes the dynamics of quarks in the gluon background field.  The
size of each vector is $\mathcal O(10^6)$.  Typically about 10~GB I/O is
carried out in production runs of length $\approx$30~minutes on
32~processors of ALiCE\@. In practice, we have to perform thousands of
such runs.

\subsection{Physical problem}
       
Non-perturbative lattice quantum chromodynamics (LQCD) deals with the
determination of hadronic properties and
interactions~\cite{Montvay:1994cy}.  Particularly important
observables are given by the mass spectrum of bound quark states, as
for instance the masses of hadrons like the $\pi$-Meson and the rho
$p$-Meson.  Among these particles, hadronic states that can be
classified as \textit{singlet representation} of the flavor-SU(3)
group play a special r\^ole.  They are characterized by contributions
of so-called non-valence objects. More precisely, their correlation
functions, $C_{\eta'}(t_1-t_2)$, the quantities which allow to extract
the physical properties of the hadrons by exploring their decay in
time, $\Delta t=t_1-t_2$, contain contributions from correlators
between closed virtual quark-gluon loops. These ``non-valence''
objects are nothing but a manifestation of quantum mechanical vacuum
fluctuations, which follow from Heisenberg's uncertainty principle as
applied to relativistic field theory.  From a physical point of view,
flavor singlet objects are particularly intriguing, as they are
sensitive to (and thus allow to explore) the topological structure of
the QCD vacuum.

The reliable determination of disconnected diagrams has been a
long-standing issue ever since the early days of LQCD\@. It can be
traced back to the numerical problem of getting information about
functionals of the complete inverse fermionic matrix
$M^{-1}$.\footnote{In contrast to flavor singlet observables,
  non-singlet masses are far simpler to compute: they imply the
  solution of a few systems of linear equations of type $Mx=b$, the
  discrete analogue of Dirac's equation with source term.}

First attempts in this direction started only a few years ago, using
the so-called stochastic estimator method (SE)~\cite{Eicker:1996gk} for the
computation of the trace of $M^{-1}$. This approach requires to solve
a linear system $Mx = \xi $ on hundreds of source vectors $\xi$, with
$\xi$ being noise vectors that are Z$_2$ of Gaussian distributed.

In Ref.~\cite{Neff:2001zr}, we have shown how to estimate the mass of
the $\eta'$ meson just using a set of low-lying eigenmodes of $M$.
Strictly speaking, our approach deals with the matrix $Q$, the
hermitian form of $M$, the eigenvectors of which form an orthogonal
base~\cite{Hip:2001hc}:
\begin{equation}
Q =\gamma_5 M.
\end{equation}
The Wilson-Dirac matrix is given by
\begin{eqnarray}
M_{\mathsf{x,y}} = \mathbf{1}_{cs}\delta_{\mathsf{x,y}}
- \kappa \sum_{\mu=1}^4 &&
{(\mathbf{1}_s-\gamma_{\mu})}\otimes U_{\mu}(\sf x)\,\delta_{\sf x,\sf
  y-\mu}\nonumber\\
+&& 
{(\mathbf{1}_s+\gamma_{\mu})}\otimes U_{\mu}^{\dagger}(\sf x-\mu)\,\delta_{\sf x,\sf y+\mu}.
\end{eqnarray}
The symbols $\gamma_{\mu}$ stand for the $4\times 4$ Dirac spin
matrices. The $3\times 3$ matrices $U_{\mu} \in$ color-SU(3) represent
the gluonic vector field, thus $\mathbf{1}_{cs}$ is a $12\times 12$
unit-matrix in color and spin space.  $M$ is a sparse matrix in
4-dimensional Euclidean space-time with matrix valued stochastically
distributed coefficient functions of type $(\mathbf{1}_s\pm
\gamma_{\mu})\otimes U_{\mu}(\mathsf{x})$ at site $\mathsf{x}$.

Once the low-lying modes are computed, it is possible to approximate
the full inverse matrix $Q^{-1}$ and those matrix functionals or
functions of $Q$ which are sensitive to small eigenvalues, i.e.\ 
long-range physics.

\subsection{Numerical procedure}
In LQCD, hadronic masses are extracted from the large-time behavior of
correlation functions. The correlator of the flavor
\textit{non-singlet} $\pi$-meson is defined as
\begin{equation}
C_{\pi} (t=t_1-t_2) = \left<
\sum_{\vec x,\vec y} \mbox{Tr}  \Big[ Q^{-1} (\vec x,t_1;\vec y,t_2) Q^{-1}
(\vec y,t_2;\vec x,t_1) \Big]  \right>_U, \label{picorr}
\end{equation}
while the flavor \textit{singlet} $\eta'$ meson correlator is composed
of two terms, 
the first 
corresponding to the propagation of a quark-antiquark-pair from $\vec
x$ to $\vec y$ without annihilation in between and the second one
being characterized by intermediate pair annihilation:
\begin{eqnarray}\label{ETAPRIME}
C_{\eta'} (t_1-t_2) &=& C_{\pi} (t_1-t_2)\nonumber\\ &-&2 \left<
\sum_{\vec x,\vec y} \mbox{Tr}  \Big[ Q^{-1} (\vec x,t_1;\vec x,t_1) \Big]
\mbox{Tr} \Big[ Q^{-1} (\vec y,t_2;\vec y,t_2) \Big]  \right>_U.
\end{eqnarray}
The brackets $\left< \ldots \right>_U$ indicate the average over a
canonical ensemble of gauge field configurations. For large time
separations $t$, the respective correlation functions are dominated by
the ground state, because the higher excitations die out and therefore
become proportional to $\exp(-m_{0}t)$. $m_0$ is the mass of the
particle associated to the correlation function.

As already mentioned, the $\pi$-correlator (\ref{picorr}), equivalent
to the first term of the $\eta'$-correlator, is obtained by solving the
linear system
\begin{equation}
M(\vec x,t_1;\vec y,t_2)\, c(\vec y,t_2) = \delta(\vec 1,1;\vec x,t_1); \label{lineq}
\end{equation}
on 12 source vectors (here located at site $(\vec 1,1)$).  Of course,
the statistics could be improved by averaging over many sources;
however this becomes prohibitively expensive as the effort increases
with the number of sources.

The second term in \eq{ETAPRIME},
\begin{equation}
\sum_{\vec x,\vec y} \mbox{Tr}  
\Big[ Q^{-1} (\vec x,t_1;\vec x,t_1) \Big] \mbox{Tr} \Big[ Q^{-1}
(\vec y,t_2;\vec y,t_2) \Big],
\end{equation}
depends on the diagonal elements of $Q^{-1}$. The inverse can be
expanded in terms of the eigenmodes weighted by the inverse
eigenvalues:
\begin{equation}
Q^{-1} (\vec x,t_1;\vec y,t_2)  = \sum_i\frac{1}{\lambda_i}
\frac{|\psi_i(\vec x,t_1)\rangle\langle\psi_i(\vec y,t_2)|}{
\langle\psi_i|\psi_i\rangle} , 
\end{equation}
where $\lambda_i$ and $\psi_i$ are the eigenvalues and the
eigenvectors of $Q$ respectively.  We found that we can approximate
the sum on the right hand side by restriction to $\mathcal O(300)$ lowest-lying
eigenvalues and their corresponding eigenvectors. Due to the factor
$1/\lambda_i$, one expects that the low-lying eigenmodes will tend to
dominate the sum. Our procedure is called truncated eigenmode
approximation (TEA).

We compute the eigensystem by means of the implicitly restarted
Arnoldi method, a generalization of the standard Lanczos procedure.  A
crucial ingredient of our approach is a Chebyshev acceleration
technique.  The spectrum is transformed such that the Arnoldi
eigenvalue determination becomes uniformly accurate for the entire
part of the spectrum we aim for.  A comfortable parallel
implementation of IRAM is provided by the PARPACK package
\cite{PARPACK}.

We work on a space-time lattice of size $16^3 \times 32$.  Taking into
account the Dirac and color indices, the Dirac matrix acts on a $12
\times 16^3 \times 32 = 1.572.864$ dimensional vector space. This
explains why the inversion of the entire Dirac matrix is not feasible
since this would need about 40~TB memory space, whereas the
determination of 300~low-lying eigenvectors leads to about 7.5~GB
memory space only. Our computations are based on canonical ensembles
of 200~field configurations with $n_f = 2$ flavors of dynamical sea
quarks. Such kind of ensembles have been generated at 5~different
dynamical quark masses in the framework of the SESAM full QCD project
\cite{Lippert:1999ha}.  It takes us about 3.5~Tflops-hours to solve
for 300~low-lying modes per ensemble for each quark mass. Altogether
we aim at $\mathcal O(30)$ valence mass/coupling combinations. First
physical results of our computations can be found in
Refs.~\cite{Neff:2001zr,Attig:2001ty,Neff:2002mq,Schilling:2001xd}.

\subsection{Benchmarks}

The typical compute partitions used for the TEA on ALiCE range from 16
to 64~nodes, depending on the number of eigenmodes required for the
approximation as well as on the memory available.  Each node reads its
specific portion of a given eigenmode corresponding to the sub-lattice
assigned to this node. In our case, a simple regular space
decomposition of the $16^3 \times 32$ lattice in $z$ and $t$
directions is applied.

Physically, the 300~eigenmodes for each field configuration are stored
as a single large file in round-robin manner. In case of a stripe-size
that corresponds to the size of an entire time-slice of the lattice,
each time-slice will be assigned to exactly one processor by PVFS\@.

In our application we use MPI-IO calls
(\texttt{MPI\_File\_read\_at()}) instead of standard I/O to
read from the PVFS\@.  In \tab{ROMIO} performance averages over four
measurements are presented.  The results fluctuate only marginally.

\begin{table}
\begin{center}
\caption{\label{ROMIO} Read and write performances for three compute partitions.}
\begin{tabular}{l|ccc}
\hline
\# of compute nodes    &   16  & 32  & 64 \\
\# of eigenmodes       &   100 & 200 & 300 \\
\hline
read performance per I/O-node [MBytes/s] &   11.9& 13.3& 11.1\\
write performance per I/O-node [MBytes/s] &  11.9& 13.4& 13.2\\
\hline
\end{tabular}
\end{center}
\end{table}

As to be expected from \fig{READ-DISK} the throughput for
read-operations is as high as 13~MB/s, which is the actual hard-disk
performance. The effective throughput for write-operations is about 10
\% less than achieved in \fig{WRITE}, most likely a remaining MPI-IO
overhead.

\subsection{Gain}

A production step includes reading eigenmodes and computing
observables. Typically the computation takes about 10~minutes. Another
10~minutes are needed for loading all 300~modes from the local disk.
Loading the data from an external archive (which is the usual
procedure since local disks do not provide enough capacity for an
entire ensemble of field configurations and eigenmodes) lasts about
30~minutes. This large mismatch between compute time and I/O time
(where the processors are idling) renders standard clusters very
inefficent for such type of computational problems, denoted as {\it
  data-intensive}.

PVFS via ParaStation/Myrinet cuts down the I/O-times to less than
20~seconds for both reading and writing. This is a substantial
acceleration compared to local or remote disk I/O.  With these
improvements we were able to enter large-scale productions with
thousands of read-compute-write sequences to be carried out. After
several months of continuous heavy duty we find the I/O system to
behave remarkably reliable and stable, with no failure encountered so
far.

\section{Future Perspectives: ClusterFile}

Most parallel file systems, including PVFS, distribute the files
stripe-wise over the I/O-nodes. However, parallel applications provide
their specific data structures that are often accessed in form of
regular patterns called the ``virtual'' or ``logical'' partitioning of
the data.  The access pattern studies~\cite{NK+96,SR97,SR98} have
shown that the performance and scalability of parallel scientific
applications with intensive parallel I/O-activity suffer significantly
from the mismatch between virtual partitioning and physical placement
of file data. This may result in an under-utilization of disk and
network bandwidths and in a decreased parallel exploitation of independent
disk capacity---among other effects.

For the above mentioned reasons, it is important that a parallel file
system offers support for flexible physical partitioning. On one hand
one would wish a fully developed PFS to be able to ``translate''
efficiently from any physical partitioning of the files on the
I/O-nodes of the PFS to any virtual partitioning and vice versa, on
the other hand one would like to control the physical data layout on
the PFS\@. So far, this goal is only partly realized by PVFS\@.
Therefore, we decided to develop a novel parallel file system, called
ClusterFile~\cite{IT01}, that addresses these issues.

In its present state, ClusterFile presents architectural similarities
with PVFS, in so far that several I/O-daemons are responsible
for storing the data on I/O-nodes with one central meta-data manager.
The clients must link to a library that provides transparent file
system access.

Unlike PVFS, ClusterFile employs a file partitioning model, that
allows the arbitrary distribution of data over a cluster, in regular
or irregular patterns. The model is optimized for regular patterns,
since most frequently used data structures of parallel scientific
applications are multidimensional arrays~\cite{NK+96}, partitioned
into chunks among parallel processes.

The file model of ClusterFile is used for both physical and logical
partitioning. A file is physically partitioned into sub-files. As an
example, it is possible to spread a file over several disks using a
block-cyclic distribution or any other regular and irregular
distribution as for instance supported by High Performance
Fortran~\cite{HPF}.

A file may be logically partitioned between several processors by
means of views. A view is a sequential window to an eventually
non-contiguous subset of a file and can be used exactly in the same
way as a file. An important advantage of using views is that it
relieves the programmer from complex index computation.  Additionally,
it also provides hints on potential future access patterns and can be
used for matching the physical to the logical distribution. For
instance, if a file is striped in round-robin manner over the disks
while the parallel processes of an application set block-cyclic views,
it could be better for performance to convert the physical layout into
a conforming block-cyclic distribution.

The processes of a parallel application access a shared file in many
cases at nearly equal times. This observation suggests the design of
collective I/O operations. Their main goal is to coalesce many small
I/O requests of several cooperating processes into a few large
requests. The two main categories of collective I/O are
two-phase~\cite{RB+93} and disk-directed~\cite{KD94} operations. The
two-phase I/O consists of a file access step that is independent of
the virtual partitioning, and a shuffle-phase in which the access data
is distributed according to the access pattern.  The MPI-IO library,
which is implemented on top of PVFS and ClusterFile utilizes the
two-phase method since it is not aware of the physical file
partitioning.  ClusterFile implements a version of the disk-directed
method, in which the requests are gathered at I/O nodes, coalesced,
before access is performed, and the result is returned to the compute
nodes. This approach has the advantage that it can exploit the
relationship between physical and logical partitionings, whereas the
two-phase method separates the operation into two distinct steps.

In the future we plan to test ClusterFile on ALiCE and to perform a
detailed comparison with PVFS\@.  We expect the collective I/O
implementation of ClusterFile to perform better than that of MPI-IO,
due to the above mentioned reasons.

Another important step will be the incorporation of cooperative caching
policies~\cite{DW94} that allow the buffer-caches of I/O- and compute-nodes
to interact. The goal is to provide a global caching policy that
provides a better utilization of buffer-caches and avoids unnecessary
disk requests.

\section{Summary}
We have demonstrated that the ParaStation3 communication system speeds
up the performance of parallel I/O on cluster computers such as ALiCE.
I/O-benchmarks with PVFS using Parastation over Myrinet achieve a
throughput for write-operations of up to 1~GB/s from a 32-processor
compute-partition, given a 32-processor PVFS I/O-partition.  These
results out-perform known benchmark results for PVFS on 1.28~Gbit
Myrinet by more than a factor of 2, a fact that is mainly due to the
superior communication features of ParaStation.  Read-performance from
buffer-cache reaches up to 2.2~GB/s, while reading from hard-disk
saturates at the cumulative hard-disk performance. The I/O-performance
achieved with PVFS using ParaStation enables us to carry out extremely
data-intensive eigenmode computations on ALiCE in the framework of
lattice quantum chromodynamics. In the future the I/O-system will be
utilized for storing and processing mass data in high energy physics
data analysis on clusters.

\section*{Acknowledgments}

This work was supported by the Deutsche Forschungsgemeinschaft as
twinning project ``Alpha-Linux-Cluster'' (Ti264/6-1 \& Li701/3-1).
Physics related work was supported under Li701/4-1 (RESH
Forschergruppe FOR 240/4-1), and by the EU Research and Training
Network HPRN-CT-2000-00145 ``Hadron Properties from Lattice QCD''.  We
thank Guido Arnold and Boris Orth for their help with the cluster
computer ALiCE\@.

\bibliographystyle{h-elsevier}
\bibliography{pvfs}

\end{document}